\documentstyle[prl,aps]{revtex}
\input epsf
\epsfverbosetrue
\def\Lp{L_{p}}

\def\rv{{\bf r}}

\def\hv{{\bf h}}
\def\nv{{\bf n}}
\def\uv{{\bf u}}
\def\vv{{\bf v}}

\def\vv{{\bf v}}

\def\Fv{{\bf F}}

\def\Iv{{\bf I}}
\def\Rv{{\bf R}}
\def\Pv{{\bf P}}

\def\kB{k}

\def\sigv{\mbox{\boldmath $\sigma$}}

\def\gamv{\mbox{\boldmath $\gamma$}}

\def\etav{\mbox{\boldmath $\eta$}}

\def\zetv{\mbox{\boldmath $\zeta$}}

\def\epsb{\bar{\epsilon}}

\def\la{\langle}
\def\ra{\rangle}

\def\ornt{\mathsf{ornt}}
\def\curv{\mathsf{curv}}
\def\tens{\mathsf{tens}}
\def\rod{\mathsf{rod}}
\def\bend{\mathsf{bend}}
\def\met{\mathsf{met}}
\def\rand{\mathsf{rand}}
\def\eq{\mathsf{eq}}

\begin{document}

\twocolumn[\hsize\textwidth\columnwidth\hsize\csname @twocolumnfalse\endcsname
\title{ Viscoelasticity of Dilute Solutions of Semiflexible Polymers }
\author{ Matteo Pasquali, V.~Shankar, and David~C.~Morse }
\address{ Department of Chemical Engineering and Materials Science, \\
         University of Minnesota, 421 Washington Ave. S.E., Minneapolis, MN 55455}
\date{submitted to Physical Review Letters, Sept. 28, 2000}
\maketitle
\begin{abstract}
We show using Brownian dynamics simulations and theory how the shear
relaxation modulus $G(t)$ of dilute solutions of relatively stiff
semiflexible polymers differs qualitatively from that of rigid rods.
For chains shorter than their persistence length, $G(t)$  exhibits 
three time regimes: At very early times, when the longitudinal 
deformation is affine, $G(t) \sim t^{-3/4}$. Over a broad intermediate 
regime, during which the chain length relaxes, $G(t) \sim t^{-5/4}$. 
At long times, $G(t)$ mimics that of rigid rods. A model of the 
polymer as an effectively extensible rod with a frequency dependent 
elastic modulus $B(\omega) \sim (i \omega)^{3/4}$ quantitatively
describes $G(t)$ throughout the first two regimes.
\end{abstract}
\vskip2pc
]

Many important biopolymers are wormlike chains with persistence 
lengths $L_p$ comparable to or larger than their contour length 
$L$. Examples are $\alpha$-helical proteins, short DNA, collagen 
fibrils, rod-like viruses, and protein filaments such as F-actin.  
The cytoskeleton of a cell is primarily a network of such polymers, 
and plays a critical role in controlling the mechanical rigidity, 
motility, and adhesion of living cells; understanding the 
viscoelastic behavior semiflexible polymers in solution is thus 
a critical problem in biophysics. Whereas the linear viscoelastic 
behavior of dilute solutions of flexible (Gaussian) and rod-like 
polymer molecules is well understood \cite{Doi}, there is thus far 
no qualitatively correct description of the viscoelasticity of 
dilute solutions of semilexible polymers over the whole range of 
frequency and time scales. Bridging the theoretical gap between 
the flexible and rigid rod limits is thus also an important open 
problem in polymer physics. Here, we present both results from 
Brownian dynamics simulations of relatively stiff semiflexible chains, 
with $\Lp \geq L$, and a simple theory that accurately describes 
their linear viscoelastic response over a very wide range of time 
scales. Both theory and simulation yield a relaxation modulus $G(t) 
\sim t^{-5/4}$ over a wide range of intermediate times, after an 
initial decay of $G(t) \propto t^{-3/4}$ at {\it very} early times, 
and before an exponential decay of $G(t)$ at long times (like that 
of rigid rods) due to diffusive tumbling of the chain orientation.

A single wormlike chain may be described by a curve $\rv(s)$, 
with a  tangent vector $\uv(s) \equiv \partial\rv(s)/\partial s$, 
where $s$ is contour distance along the chain.  Inextensibility 
requires that $|\partial\rv/\partial s|=1$.  The bending energy 
of a chain with rigidity $\kappa$ or persistence length $\Lp 
\equiv \kappa/\kB T$ is 
$U = \case{1}{2} \kappa \int\!ds\;|\frac{\partial\uv(s)}{\partial s}|^{2}$. 
The Brownian motion of such a chain in a homogenous flow 
$\vv(\rv,t) \equiv \dot{\gamv}(t) \cdot \rv $ may be described 
in a free-draining approximation \cite{Note} by a Langevin equation
\begin{equation}
  \zeta \left \{ 
  \frac{\partial \rv}{\partial t} - \dot{\gamv}\cdot\rv \right \}
   = -\kappa \frac{\partial^{4}\rv}{\partial s^{4}}
     + \frac{\partial (\uv {\cal T}) }{\partial s}
     + \etav \quad. \label{Langevin}
\end{equation}
Here ${\cal T}$ is a tension that acts to impose the constraint 
$|\partial \rv/\partial s|=1$, $\zeta$ is a friction coefficient,
and $\etav$ is a Brownian force with correlations $\la\etav(s,t)
\etav(s',t') \ra = 2 \kB T \zeta \Iv \delta(t-t') \delta(s-s')$. 
This equation can be made dimensionless in terms of reduced variables 
$\hat{t}=t \kB T/(\zeta L^{3})$, $\hat{s}=s/L$, $\hat{\rv}=\rv/L$, 
and $\hat{\Lp}=\Lp/L$.
The linear viscoelasticity of a solution of worm-like chains may 
be characterized by either the shear relaxation modulus $G(t)$, 
which describes the stress $\sigv (t) = G(t) [\gamv + \gamv^{T}]$ 
at time $t$ after an infinitesimal step strain $\gamv$, or, 
equivalently, by the complex modulus 
$G^{*}(\omega)\equiv i\omega\int_{0}^{\infty}\!dt\;G(t)e^{-i\omega t}$, 
which describes the response to a small oscillatory strain.
The polymer contribution to the moduli per chain, in a dilute 
solution of $c$ chains per unit volume in a solvent of viscosity 
$\eta_{s}$ is given by a corresponding intrinsic moduli 
$[G(t)] \equiv [G(t) - \eta_{s}\delta(t)]/c$ and 
$[G^{*}(\omega)] \equiv [G^{*}(\omega) - i\omega \eta_{s}]/c$. 
For worm-like chains, $[G(t)]$ must have the form $[G(t)]= \kB T 
\hat{G}(\hat{t},\hat{\Lp})$.

Prior work has identified some relevant time scales and provided 
predictions for $G(t)$ in several limits:

Rod-like chains ($L \ll \Lp$) should behave like rigid rods at 
$t \agt \tau_{\perp}$, where $\tau_{\perp} \equiv \zeta 
L^{4}/(\kB T \Lp)$ is roughly the relaxation time of the longest 
wavelength bending mode. The predicted modulus for dilute rigid 
rods \cite{Doi,Auer} is
\begin{equation}
   \lim_{\Lp \rightarrow \infty}
   [G(t)] = \frac{\zeta L^{3}}{180}\delta(t)
         + \frac{3 \kB T}{5} e^{-t/\tau_{\rod}} \quad,
   \label{Grod}
\end{equation}
where $\tau_{\rod} \equiv \zeta L^{3}/(72 \kB T)$ is a rotational 
diffusion time.  The exponential contribution to $[G(t)]$ is due 
to an entropic orientational stress caused by an anisotropic 
distribution of rod orientations; it decays by rotational diffusion. 
The delta-function contribution arises from the longitudinal 
tension induced in the rods during the step deformation; it decays 
instantaneously after the deformation.

In Refs.\cite{Gittes98,Morse98}, the authors considered how this
behavior is modified by the longitudinal compliance of a semiflexible 
chain. They calculated the magnitude of changes in the end-to-end 
length of a worm-like chain due to changes in the magnitude of 
transverse fluctuations when the chain is subjected to an oscillatory 
tension at frequency $\omega$, and showed that ratio of tension to
strain is given by a frequency-dependent effective longitudinal modulus 
\begin{equation}
   B(\omega) = \frac{2^{3/4}\kB T}{\Lp} ( i\omega\tau_{p})^{3/4} 
   \label{Bomega} \quad.
\end{equation}
at all $\omega \gg \tau_{\perp}^{-1}$, where $\tau_{p} \equiv 
\zeta \Lp^{3}/\kB T$. They also predicted a macroscopic viscoelastic
modulus $[G^{*}(\omega)] = LB(\omega)/15 \propto (i\omega )^{3/4}$ 
\cite{Gittes98,Morse98} at very high frequencies, or, equivalently, 
$[G(t)] \propto t^{-3/4}$ at early times, by assuming that that the
frictional coupling between the chain and the solvent must become
strong enough at very high frequencies to produce an affine 
longitudinal strain. In Refs. \cite{Morse98,Everaers98}, the authors 
considered the dynamics of longitudinal relaxation. They showed that 
longitudinal strain propagates along a chain by an anomolous diffusion 
with a frequency-dependent diffusivity $D(\omega) = B(\omega) / 
\zeta_{\parallel}$, in which the strain diffuses a distance 
$\xi_{\parallel}(t) \propto \sqrt{D(\omega = 1/t)t} \propto t^{1/8}$ 
in time $t$. Both the assumption of affine deformation and the 
predicted $t^{-3/4}$ decay of $G(t)$ must thus fail beyond the 
time $\tau_{\parallel} \equiv \zeta L^{8}/(\kB T\Lp^{5})$ required 
for strain to diffuse the chain length $L$, and so allow signficant
longitudinal relaxation. 

This prior work does not predict the behavior of $G(t)$ for rod-like 
chains over a wide range of at intermediate times $\tau_{\parallel} 
< t < \tau_{\perp}$, where relaxation of chain length and transverse 
fluctuations must be coupled. This interval must rapidly broaden as 
$L \ll \Lp$ because $\tau_{\parallel}/\tau_{\perp} \propto (L/\Lp)^{4}$. 
For $L \sim \Lp$, the gaps between $\tau_{\parallel}$, $\tau_{\perp}$, 
and $\tau_{\rod}$ vanish, and so the intermediate regime must 
disappear. Coil-like chains ($L \gg \Lp$) are expected \cite{Morse98} 
to crossover smoothly from $G(t) \propto t^{-3/4}$ to Rouse-like 
behavior $G(t) \propto t^{-1/2}$ at $t \sim \tau_{p}$, which is roughly 
the relaxation time of a bending mode of wavelength $\Lp$.

Our simulations use discrete worm-like chains of $N$ beads at 
positions $\Rv_{1}, \ldots, \Rv_{N}$ connected by $N-1$ rods of fixed 
length $a$, with unit tangents $\uv_{i} = (\Rv_{i+1}-\Rv_{i})/a$, and 
a bending energy $U=-(\kappa/a)\sum_{i=2}^{N-1}\uv_{i}\cdot\uv_{i-1}$. 
We use a midstep algorithm \cite{Hinch96} to compute bead trajectories 
generated by the equation of motion
\begin{equation}
  \zeta_b \left \{ \frac{d\Rv_{i}}{dt} \! - \! \dot{\gamv}\cdot\Rv_{i} \right \}
  \! = \! \Fv_i
  \! = \! \Fv_{i}^{\bend} 
  \! + \! \Fv_{i}^{\met} 
  \! + \! \Fv_{i}^{\tens} 
  \! + \! \Fv_{i}^{\rand}
  \; . \label{Langevin-discrete}
\end{equation}
Here, $\zeta_{b} = \zeta a$ is a bead friction coefficient,
$\Fv_{i}^{\bend} \equiv -\partial U/\partial \Rv_{i}$ is a 
bending force, $\Fv_{i}^{\rand}$ is a Langevin noise, and 
$\Fv^{\tens}_{i} = {\cal T}_{i}\uv_{i} - {\cal T}_{i-1}\uv_{i-1}$ 
is a constraint force, where ${\cal T}_{i}$ is the tension in rod 
$i$. The tensions are computed by solving $\sum_{j=1}^{N-1} H_{ij}
{\cal T}_{j} = \uv_{i} \cdot ( \tilde{\Fv}_{i+1} -\tilde{\Fv}_{i})$, 
where $\tilde{\Fv}_i \equiv \Fv_{i}^{\bend} + \Fv_{i}^{\met} 
+ \Fv_{i}^{\rand} +\zeta_{b}\dot{\gamv}\cdot\Rv_{i}$, and
$H_{ij}$ is a tridiagonal matrix with $H_{ii} = 2$ and 
$H_{ij} = - \uv_{i} \cdot\uv_{j}$ for $i=j \pm 1$.
$\Fv_{i}^{\met} \equiv \kB T \frac{\partial}{\partial \Rv_{i}}
\ln \sqrt{\det(H)}$ is a ``metric'' force that must be included 
in simulations with constrained rod lengths to obtain a Boltzmann 
distribution $e^{-U(\uv_{1},\ldots,\uv_{N-1})/\kB T}$ of rod 
orientations in thermal equilibrium \cite{Hinch96,Fixman78}. 

The modulus $[G(t)]$ is obtained from equilibrium simulations
($\dot{\gamv}=0$) by evaluating the Green-Kubo relation $[G(t)] 
= \la\sigma_{xy}(t)\sigma_{xy}(0) \ra / \kB T$, where $\sigv 
\equiv - \sum_{i} \Rv_{i} \Fv_{i} $ is the single-chain stress 
tensor. The Brownian contribution to $\sigv(t)$ is computed by 
the method of Refs. \cite{Hinch96,Doyle97}.  A wide range of 
timescales is explored at each value of $L/L_p \equiv N a /L_p$ 
by using coarser- and finer-grained chains to resolve longer and 
shorter times, respectively. Results for chains with the same 
$L/\Lp$ but different $N$ are collapsed onto master curves of 
$[G(t)]$ versus $t/\tau_{\rod}$, where $\tau_{\rod} = \zeta_{b}
a^{2}N^{3}/72 \kB T$.  Because initial chain conformations are 
chosen from a Boltzmann distribution, behavior at short times 
can be obtained from short simulations of fine-grained chains 
\cite{Everaers98}.

To elucidate the physical origins of stress, we decompose $\sigv$ as 
a sum $\sigv = \sigv_{\ornt}+\sigv_{\curv}+\sigv_{\tens} - \kB T\Iv$ 
of the orientation, curvature, and tension stresses 
\cite{Morse98}, where
\begin{eqnarray}
   \sigv_{\ornt} & \equiv & 
   \case{3}{2}\kB T( \uv_{1}\uv_{1}+\uv_{N-1}\uv_{N-1} - \case{2}{3}\Iv )
   \nonumber \\
   \sigv_{\curv} & \equiv & 
   \! - \! \sum_{i=1}^{N} \Rv_{i} \Fv_{i}^{\bend}
   \! + \! 3 \kB T\sum_{i=1}^{N-1} (\uv_{i}\uv_{i} 
   \! - \! \case{1}{3}\Iv) 
   \! - \! \sigv_{\ornt} 
   \label{sig-oct} \; .
\end{eqnarray}
and $\sigv_{\tens} = \sigv - \sigv_{\ornt} - \sigv_{\curv} + \kB T\Iv$. 
We also decompose $[G(t)]$ as $[G(t)]=[G_{\ornt}(t)]+[G_{\curv}(t)]
+[G_{\tens}(t)]$, where $[G_{\alpha}(t)] = \la \sigma_{\alpha,xy}(t)
\sigma_{xy}(0) \ra / \kB T$, with $\alpha=$ ``$\ornt$'', ``$\curv$'', or 
``$\tens$'', describes the decay of the stress $\la \sigv_{\alpha}(t) \ra$ 
after a hypothetical step deformation. $\sigv_{\curv}$ arises from 
disturbances of the equilibrium distribution of bending mode fluctuations, 
and was predicted to vanish for rod-like chains at times $t \agt \tau_{\perp}$;
$\sigv_{\ornt}$ is an analog of the orientational stress of a solution 
of rigid rods; and $\sigv_{\tens}$ is the stress arising from longitudinal 
tension \cite{Morse98}.

\begin{figure}[t]
\centerline{\epsfxsize=7.75cm \epsfbox{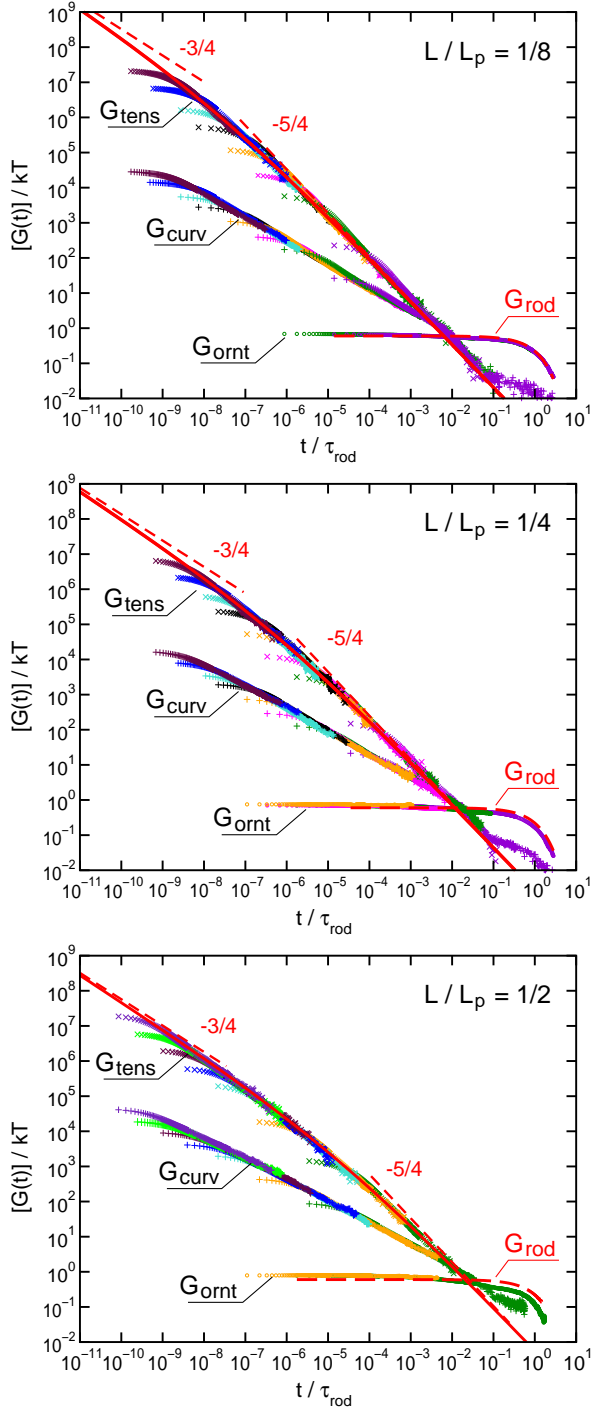}}
\vspace*{2mm}
\caption{
Simulation results for $[G_{\tens}(t)]$ (top curve in each plot,
$\times$), $[G_{\curv}(t)]$ (middle curve, $+$), and $[G_{\ornt}(t)]$
(bottom curve, $\circ$) {\it vs}. $t/\tau_{\rod}$, with
$\tau_{\rod}=\zeta_{b}N^{3}a^{2}/(72 \kB T)$, for:
$L/L_{p}=1/8$ and $N=8,16,22,32,46,64,90,128$ (top plot)
$L/L_{p}=1/4$ and $N=8,16,22,32,46,64,90,128$ (middle plot), and
$L/L_{p}=1/2$ and $N=16,32,64,90,128,180,256$ (bottom plot). In
each plot, $[G_{\ornt}(t)]$ is shown only for a few small values 
of $N$. The long-dashed red line is the prediction $[G_{\rod}(t)] =
\frac{3}{5} \kB T e^{-t/\tau_{\rod}}$ for rigid rods at $t > 0$.
Short-dashed red lines with slopes of $-3/4$ and $-5/4$ are 
asymptotes Eqs.~(\ref{Gearly}) and (\ref{Ginter}).  The solid red 
line is the predicted $[G_{\tens}(t)]$ obtained by Fourier 
transforming Eq.~(\ref{Gtheory}). }
\label{Fig1} 
\end{figure}

Fig.~(\ref{Fig1}) shows master curves of $[G_{\tens}(t)]$, 
$[G_{\curv}(t)]$, and $[G_{\ornt}(t)]$ for chains of reduced length 
$L/\Lp=1/8$, $1/4$, $1/2$.  The regions of overlap of results obtained 
with different values of $N$ (which are shown by different colors), 
reflect the behavior of a continuous chain, while the saturation of 
$[G_{\tens}(t)]$ and $[G_{\curv}(t)]$ to $N$-dependent limiting values 
at small $t$ is due to the discreteness of the chains. At long times, 
$t \agt \tau_{\perp}$, the largest contribution to $[G(t)]$ is 
$[G_{\ornt}(t)]$, which approaches the exponential relaxation predicted 
for a rigid rod solution. At $t 
\sim \tau_{\perp}$, all three contributions to $[G(t)]$ are comparable.  
At earlier times, $[G(t)]$ is dominated by $[G_{\tens}(t)]$.  For the 
most flexible chains shown ($L=\Lp/2)$, $[G_{\tens}(t)]$ closely 
approaches the predicted $t^{-3/4}$ asymptote at small $t$. For the 
two stiffer systems, $[G_{\tens}(t)]$ does not reach this asymptote
within the accessible range of $t$, and decays more rapidly than 
$t^{-3/4}$ in this range.

These results are consistent with the prediction of a $t^{-3/4}$ decay 
of $[G_{\tens}(t)]$ below a reduced time $\tau_{\parallel} /\tau_{\rod} 
\propto (L/\Lp)^{5}$ that drops rapidly as $L/\Lp$ decreases, and
suggest the possible existence of a second power law in the intermediate 
time regime $\tau_{\parallel} \ll t \ll \tau_{\perp}$ for $L \ll \Lp$. 
By postulating the existence of an intermediate power law that meets 
the predicted $t^{-3/4}$ asymptote at $t \sim \tau_{\parallel}$, and 
that falls to $[G_{\tens}(t)] \sim \kB T$ at $t \sim \tau_{\perp}$, 
we obtain $[G_{\tens}(t)] \sim \kB T (t/\tau_{\perp})^{-5/4}$. The
exponent $-5/4$ agrees with the observed slope of $\log [G_{\tens}(t)]$ 
vs.~$\log(t)$ for the
stiffest systems shown ($L=\Lp/8$), which displays the widest 
intermediate time regime.

We now present a theory of the longitudinal dynamics of a rod-like chain 
that predicts the observed $t^{-5/4}$ decay of $[G_{\tens}(t)]$ at 
intermediate times for $L\ll\Lp$.  Our analysis resembles one given 
previously to describe longitudinal relaxation of entangled chains 
\cite{Morse98}. For rod-like chains, we may expand $\rv(s,t)$ around 
a rod-like reference state as $\rv(s,t) = r_{\parallel}(s,t)\nv(t) + 
\hv(s,t)$, where $\hv(s,t)$ satisfies $\hv(s,t) \cdot \nv(t)=0$, and 
$\nv(t)$ is a unit vector that rotates with the flow like a non-Brownian 
rigid rod: $\dot{\nv} = \Pv\cdot \dot{\gamv} \cdot\nv$, where $\Pv 
\equiv \Iv - \nv\nv$. Linearizing Eq.~(\ref{Langevin}) then yields 
longitudinal and transverse equations 
\begin{eqnarray}
  \zeta
  \{ \frac{\partial r_{\parallel}}{\partial t} 
  - r_{\parallel}\dot{\gamv}:\nv\nv \}
  & = & \frac{\partial {\cal T}}{\partial s} + \eta_{\parallel} 
  \label{fLangevin} \\
  \zeta
  \Pv\cdot\{ \frac{\partial \hv}{\partial t} - \dot{\gamv}\cdot\hv \}
  & = & - \kappa\frac{\partial^{4}\hv}{\partial s^{4}} + 
  \frac{\partial }{\partial s}\left ( 
  {\cal T}\frac{\partial \hv}{\partial s} \right )
  + \etav_{\perp}.
  \label{hLangevin} 
\end{eqnarray}
These equations are coupled by the tension ${\cal T}$, which is chosen 
to satisfy the constraint $|\frac{\partial\rv}{\partial s}|^{2}= 
|\frac{\partial r_{\parallel}}{\partial s}|^{2}
+|\frac{\partial \hv}{\partial s}|^{2}=1$. 

It is convenient to introduce a longitudinal strain field 
$\epsilon(s) \equiv  \frac{\partial r_{\parallel}(s)}{\partial s}
- \la \frac{\partial r_{\parallel}}{\partial s} \ra_{\eq}$, where 
$\la \cdots \ra_{\eq}$ denotes a thermal equilibrium average. 
Combining this definition with the constraint and expanding to 
lowest order in $|\partial \hv/ \partial s|^{2}$ yields an 
approximate expression of $\epsilon(s)$ in terms of $\hv(s)$,
$  \epsilon(s)  \simeq - \frac{1}{2} \left \{
   | \frac{\partial\hv(s)}{\partial s} |^{2}  -
  \la | \frac{\partial\hv(s)}{\partial s} |^{2} \ra_{\eq} 
  \right \}  $.
This expression for $\epsilon(s)$ and Eq.~(\ref{hLangevin}) were
used in Refs \cite{Gittes98,Morse98} to calculate the linear 
response of the spatial average strain $\la \epsb(\omega) \ra 
\equiv \int\!ds \la \epsilon(s,\omega)\ra/L$ to a spatially uniform 
oscillating tension ${\cal T}(\omega)$ at frequency $\omega$
(where functions of $\omega$ denote Fourier amplitudes), yielding 
an effective extension modulus $B(\omega) \equiv {\cal T}(\omega)
/\la \epsb(\omega) \ra$ that is given by Eq.~(\ref{Bomega}) at 
$\omega \gg \tau_{\perp}^{-1}$,\cite{Gittes98,Morse98} and by 
a static value $B(0) \sim \kB T\Lp^{2}/ L^{3}$ at $\omega \ll 
\tau_{\perp}^{-1}$.\cite{MacKintosh95}

A modified diffusion equation for the strain field may be obtained 
by taking the thermal average of Eq.~(\ref{fLangevin}), differentiating 
with respect to $s$, Fourier transforming with respect to $t$, and 
setting $\la{\cal T} (s,\omega)\ra = B(\omega) \la \epsilon(s,\omega) 
\ra$. This yields
\begin{equation}
  \left ( i\omega  - \frac{B(\omega)}{\zeta}
  \frac{\partial^{2}}{\partial s^{2}} \right )
  \la \epsilon(s,\omega) \ra  \simeq 
  i\omega \gamv(\omega):\nv\nv
  \label{diffuse}
\end{equation}
where $B(\omega)/\zeta$ is an effective diffusivity and 
$\gamv(\omega)$ is the amplitude of an oscillatory strain tensor. 
Hereafter $\nv(t)$ is approximated by its time average over one 
period of oscillation. Eq.~(\ref{diffuse}), with 
$\la\epsilon(s,\omega)\ra=0$ at the chain ends, has the solution 
\begin{equation}
  \la\epsilon(s,\omega)\ra= \left [ 1-
  \frac{\cosh(\lambda(\omega)(2\hat{s}-1))}
  {\cosh(\lambda(\omega))} \right ] \gamv(\omega):\nv\nv
  \quad, \label{etheory}
\end{equation}
where $\lambda(\omega) \equiv (i\omega\zeta L^{2}/4B(\omega))^{1/2}
= (i\omega\tau_{\parallel}/2^{11})^{1/8}$.

The tension stress of rod-like chains subjected to an 
infinitesimal oscillatory strain $\gamv(\omega)$ is given by
\begin{equation}
   \sigv_{\tens}(\omega) \simeq
   \int_{0}^{L}\!ds\;\la{\cal T}(s,\omega)\
   \nv\nv \ra \label{sigtensint} \quad.
   \label{sigmatens}
\end{equation}
where $\la\cdots \ra$ denotes an average over both weak 
fluctuations and overall rod orientations. 
Combining Eq.~(\ref{sigmatens}) with Eq.~(\ref{etheory})
for the strain along a rod of orientation $\nv$ and 
averaging over random rod orientations yields a stress 
$\sigv_{\tens}(\omega) = [G^{*}_{\tens} (\omega)] 
[\gamv(\omega)+\gamv^{T}(\omega)]$, with a modulus
\begin{equation}
   [G^{*}_{\tens}(\omega)] = \case{1}{15}LB(\omega)
   \left [ 1 - \frac{ \tanh( \lambda(\omega) ) }
   { \lambda(\omega) } \right ]
   \quad. 
   \label{Gtheory}
\end{equation}
This prediction has the following limiting behaviors: 
At frequencies $\omega \gg \tau_{\parallel}^{-1}$, 
Eq.~(\ref{Gtheory}) reduces to the high-frequency asymptote
$[G_{\tens}^{*}(\omega)] \simeq L B(\omega)/15$ found previously
\cite{Gittes98,Morse98}.
Fourier transforming this asymptote yields a relaxation modulus
\begin{equation}
  \lim_{t \ll \tau_{\parallel} }
  [G_{\tens}(t)] = C_{1} \frac{\kB TL}{\Lp} 
  \left ( \frac{t}{\tau_{p}} \right )^{-3/4}
  \quad, \label{Gearly}
\end{equation}
where $C_{1}=2^{3/4}/[15\Gamma(\case{1}{4})]=0.0309$. At 
intermediate frequencies $\tau_{\parallel}^{-1} \gg \omega \gg 
\tau_{\perp}^{-1}$, where $\lambda(\omega) \ll 1$, expanding 
Eq.~(\ref{Gtheory}) in powers of $\lambda(\omega)$ yields a 
modulus 
$[G^{*}_{\tens}(\omega)] = i\omega \frac{\zeta L^{3}}{180} 
- \frac{\kB T}{1800 \; 2^{3/4}}(i\omega\tau_{\perp} )^{5/4}+\cdots$,
which includes a dominant contribution of order $i\omega$, whose
prefactor is identical to that found for rigid rods, and a first 
correction proportional to $(i\omega)^{5/4}$.  This yields a 
loss modulus $[G''(\omega)] \propto \omega$ (like rigid rods) 
but a storage modulus $[G'(\omega)] \propto \omega^{5/4}$ 
(unlike rigid rods) at these frequencies. Upon transforming this 
intermediate asymptote, the term proportional to $i\omega$ yields 
an apparent $\delta$-function contribution to $[G(t)]$ (as for 
rigid rods), and so $[G_{\tens}(t)]$ is instead dominated at 
$\tau_{\parallel} \ll t \ll \tau_{\perp}$ by the transform of 
the term of proportional to $(i\omega)^{5/4}$, which yields 
\begin{equation}
  \lim_{\tau_{\parallel} \ll t \ll \tau_{\perp}}
  [G_{\tens}(t)] \simeq  C_{2} \kB T
  \left ( \frac{t}{\tau_{\perp}} \right )^{-5/4}
  \quad, \label{Ginter}
\end{equation}
where $C_{2} = 1/[2^{3/4} 7200\Gamma(\case{3}{4})] = 0.0000674$.
At $\omega \alt \tau_{\perp}^{-1}$ or $t \agt \tau_{\perp}$, 
Eq.~(\ref{Bomega}) for $B(\omega)$ becomes inapplicable, but 
$[G_{\tens}(t)]$ also becomes small compared to $[G_{\ornt}(t)]$.

The predictions of $[G_{\tens}(t)]$ shown in Fig. (\ref{Fig1}) 
were obtained by Fourier transforming Eq.~(\ref{Gtheory}) 
numerically.  They agree with the simulation results for 
$[G_{\tens}(t)]$ at all $t \alt \tau_{\perp}$, and accurately
describe not just the power law regimes, but the broad 
crossovers between them. Remarkably, the theory remains 
accurate for $L = \Lp/2$, despite the assumption of a nearly 
straight chain.

Our derivation of Eq.~(\ref{diffuse}) explicitly assumes a  {\it
local} proportionality of the tension and strain at  each point on 
the chain, with $\la\epsilon(s,\omega)\ra  = B^{-1}(\omega)\la {\cal
T}(s,\omega)\ra$, rather than  allowing for a spatially nonlocal
response of the form $\la\epsilon(s,\omega)\ra=\int\!ds'
\;B^{-1}(s,s',\omega)  \la{\cal T}(s',\omega)\ra$. To examine this
approximation, we calculated the nonlocal compliance $B^{-1}(s,s',
\omega)$.  We find that the range of nonlocality is of the order of 
the wavelength $\xi_{\perp}(\omega) = (\omega \zeta/\kB T\Lp)^{-1/4}$ 
of the bending mode with frequency $\omega$, 
and that the strain predicted by Eq.~(\ref{diffuse}) varies slowly 
over lengths of order $\xi_{\perp}(\omega)$ for all $\omega \agt 
\tau_{\perp}^{-1}$. This justifies our local compliance approximation 
for all $\omega \agt \tau_{\perp}^{-1}$.

A conceptually simple, analytically solvable, and accurate model 
of the dominant contribution to $[G(t)]$ at times $t \alt 
\tau_{\perp}$, valid for all $L \alt \Lp$, is thus obtained by 
treating the inextensible worm-like chain as an effectively 
extensible rod with a frequency-dependent longitudinal modulus 
given by Eq.~(\ref{Bomega}). At later times, $[G(t)]$ is 
dominated by $[G_{\ornt}(t)]$, which mimics the behavior of a 
solution of rods. The simulations show that the curvature stress
never dominates $[G(t)]$ in such solutions.  A useful global 
approximation for $[G(t)]$ for rod-like chains may thus be 
obtained simply by replacing the $\delta$-function in Eq.~(\ref{Grod}) 
by our result for $[G_{\tens}(t)]$. Our results confirm that 
$[G(t)]$ initially decays as $t^{-3/4}$, but also show that, 
when $L \alt \Lp$, this behavior is observable only below a 
time proportional to $\tau_{\parallel}$ that drops rapidly 
with decreasing $L/\Lp$ to values inaccessible to either 
simulation or experiment.  Therefore, measurements of the 
viscoelastic modulus of dilute solutions of rod-like chains 
at practically attainable high-frequencies may often probe either 
the $t^{-5/4}$ regime identified here, instead of the initial 
$t^{-3/4}$ regime, or the broad---but calculable---crossover 
between them.

{\it Acknowledgements}: This work was supported by 
NSF DMR-9973976 and Minnesota Supercomputing Institute.

\end{document}